%% file: ms.v2.tex
\documentclass[preprint2]{aastex}                
\newcommand{\funits}{erg/s/cm$^2$/{\AA}}
\newcommand{\fuse}{{\it FUSE}}
\newcommand{\fcont}{$F_{\lambda1060}$}
\newcommand{\hd}{HD\,5980}
\newcommand{\hst}{{\it HST}}
\newcommand{\iue}{{\it IUE}}
\newcommand{\kms}{km\,s$^{-1}$}
\newcommand{\lsun}{$L_\sun$}
\newcommand{\mdot}{$\dot{M}$}
\newcommand{\msun}{${\rm M}_\sun$}
\newcommand{\msunpyr}{${\rm M}_\sun / {\rm yr}$}
\newcommand{\orfeus}{{\it ORFEUS}}
\newcommand{\rsun}{$R_\sun$}
\newcommand{\stA}{star A}
\newcommand{\stB}{star B}
\newcommand{\stC}{star C}
\newcommand{\vinf}{${v}_\infty$}

\newcommand{\vturb}{${v}_{turb}$}
\slugcomment{Version of \today}                                                %
\shortauthors{Koenigsberger et al.}                                            %
\shorttitle{The Eruptor's Wind in {\hd}}                                       %
\begin{document}

\title{{\fuse} Observations of {\hd}: The Wind Structure of the Eruptor\footnote{Based on 
observations made with the NASA-CNES-CSA Far Ultraviolet Spectroscopic Explorer.
{\fuse} is operated for NASA by the Johns Hopkins University under NASA
Contract NAS5-32985.} }

\author{Gloria Koenigsberger}
\affil{Centro de Ciencias Fisicas, 
       Universidad Nacional Aut\'{o}noma de M\'{e}xico,
       {Adpo. Postal 48-3}, 
       Cuernavaca, Morelos
       62251 Mexico}
\email{gloria@fis.unam.mx}

\author{Alexander W. Fullerton\altaffilmark{2}}
\affil{Dept. of Physics \& Astronomy,
       University of Victoria,
       P.O. Box 3055,
       Victoria, BC, {V8W 3P6}, Canada.}
\altaffiltext{2}{Postal Address: 
                 Dept. of Physics \& Astronomy,
                 The Johns Hopkins University,
		 3400 N. Charles Street,
		 Baltimore, MD 21218.}
\email{awf@pha.jhu.edu}

\author{Derck Massa}
\affil{SGT, Inc,
       NASA's Goddard Space Flight Center,
       Code 681.0,
       Greenbelt, MD 20771.}
\email{massa@taotaomona.gsfc.nasa.gov}

\and 

\author{Lawrence H. Auer}
\affil{Earth and Environmental Sciences 5,
       Los Alamos National Laboratory,
       MS-F665,
       {Los Alamos}, NM 87545}
\email{lhainnm@mindspring.com}
\begin{abstract}
{\hd} is a unique system containing one massive star ({\stA}) that is 
apparently entering the luminous blue variable phase, and an eclipsing 
companion ({\stB}) that may have already evolved beyond this phase to 
become a Wolf-Rayet star. 
In this paper we present the results from {\fuse} observations obtained in 
1999, 2000, and 2002 and one far-UV observation obtained by {\orfeus}/BEFS in 
1993 shortly before the first eruption of {\hd}.   
The eight phase-resolved spectra obtained by {\fuse} in 2002 are analyzed in 
the context of a wind-eclipse model.
This analysis shows that the wind of the eruptor obeyed a very fast velocity law 
in 2002, which is consistent with the line-driving mechanism.
Large amplitude line-profile variations on the orbital period are shown
to be due to the eclipse of {\stB} by the wind of {\stA}, although the 
eclipse due to gas flowing in the direction of {\stB} is absent.  
This can only be explained if the wind of {\stA} is not spherically 
symmetric, or if the eclipsed line radiation is ``filled-in" by emission 
originating from somewhere else in the system, e.g., in the wind-wind collision 
region.  
Except for a slightly lower wind speed, the {\orfeus}/BEFS spectrum is very 
similar to the spectrum obtained by {\fuse} at the same orbital phase: there is
no indication of the impending eruption. 
However, the trend for decreasing wind velocity suggests the occurrence of
the ``bi-stability" mechanism, which in turn implies that the restructuring 
of the circumbinary environment caused by the transition from 
``fast, rarefied wind" to ``slow, dense wind" was observed as the 
eruptive event. 
The underlying mechanism responsible for the long-term decrease in wind
velocity that precipitated this change remains an open issue.
\end{abstract}

\section{Introduction}\label{intro}

{\hd} (Sk\,78, AV\,229) is the most luminous eclipsing binary system in the
Small Magellanic Cloud (SMC) and a member of the young stellar cluster
NGC\,346.
It is an ideal object for studying interacting winds and the 
evolution of very massive stars in binaries.  
The system achieved notoriety in 1994, when its primary star 
(hereafter {\stA}; M$\sim$50 \msun) was discovered to be undergoing 
an eruptive event of a yet-undetermined nature.  
Although the system had been monitored since the early 1980s, the sudden 
brightening by $\sim$2 magnitudes in late 1993 went unrecorded except for 
the visual estimates of A. Jones \citep{Bateson93}.  
An intensive ultraviolet (UV) monitoring campaign was initiated in
mid-1994, and this database provides a unique record of the
declining phase of the outburst.
However, the most intriguing question that {\hd} raises involves 
the mechanism responsible for the eruption.
We know that between 1987 and 1995 its UV spectrum went successively 
through all the ``late" nitrogen Wolf-Rayet (WR) classes --
WN6 $\rightarrow$ WN7 $\rightarrow$ WN8 $\rightarrow$ WN11 $\rightarrow$ WN6 -- 
while also exhibiting significant changes in wind velocity and visual brightness;
see \citet{Koenigsberger04} for a general review.    
Although some of the characteristics of the eruption are similar to those 
observed in Luminous Blue Variables (LBVs), both its extraordinary luminosity 
($\sim3 \times 10^6$~{\lsun}) and its increase in luminosity during 
eruption \citep[$\sim10^7$~\lsun;~][]{Drissen01} place it in a category
currently shared only with $\eta$ Carinae.  

The second component of the eclipsing system (hereafter {\stB}; 
M$\sim$28 \msun) also appears to be a WR star.  
It has been difficult to isolate its spectral characteristics from the combined 
spectrum; see \citet{Koenigsberger04} for a discussion of this issue,
which is further complicated by the possibility that some emission at line 
frequencies arises in a wind-wind colliding region \citep{Moffat98}.  
If {\stB} is indeed a WNE star, then it is the evolved remnant of the 
originally more massive star of the binary system, and is therefore 
a good pre-supernova candidate.
Alternately, it could simply be a less massive star whose outer layers
have been significantly modified by a mass-transfer process, which has led 
to the current state of {\stA}.  


The spectrum of {\hd} indicates the presence of a third star.
It can be seen in the continuum \citep{Breysacher91}, in the optical 
spectra obtained in the 1980s \citep{Niemela88}, and in the 
``stationary" UV photospheric lines that are visible in {\it HST/STIS} 
spectra \citep{Koenigsberger02}.
This component (hereafter {\stC}) is likely to be an O4--6 supergiant 
that could be gravitationally bound to the {\stA}$+${\stB} pair with a 
very long orbital period.
Thus, it might be responsible for the instability of the system 
\citep{Koenigsberger94}.
Alternately, {\stC} could simply be a line-of-sight object  
whose light contaminates the spectrum of the {\hd} system. 
The presence of this third star complicates the interpretation of an 
already complicated system.
However, because the evolution of massive stars plays such an 
important role in many astrophysical phenomena, it is important to understand 
the mechanisms responsible for the peculiar behavior of {\hd}.


In this paper, we analyze a set of phase-resolved spectroscopic observations 
of {\hd} obtained with the {\it Far Ultraviolet Spectroscopic Explorer} (\fuse) 
satellite.
These data were obtained over an interval spanning many orbital cycles,
but include a subset of 8 observations taken over $\sim$6 orbits, which permit 
orbital changes in the structure of the wind of {\stA} to be assessed.
In addition, we use an archival spectrum obtained by {\orfeus} shortly 
before {\hd} erupted.
This fortuitous observation constrains the time scale for the development
of the instability leading to the eruptions.
Our analysis is organized as follows.
In \S\ref{obs}, we describe the observational material.
The {\fuse} light curve is presented in \S\ref{fuselc}, while the
far-UV spectrum of {\hd} and its orbital variability are discussed 
in \S\ref{fusesp}.
In \S\ref{model} we analyze the orbital variations in terms of a 
wind-eclipse model, and discuss the implications.
Additional discussion and our conclusions are presented in \S\ref{concl}.

\section{Observations and Data Analysis}\label{obs}

\subsection{Observational Material}

{\hd} has been observed several times with {\fuse}, though only
those observations obtained during program P223 (PI: D.~Massa) provide 
phase-resolved coverage.
Program P223 produced 8 observations that are well distributed in orbital phase,
which were obtained over an interval of $\sim$116 days (i.e., $\sim$6
consecutive orbital cycles) in 2002. 
Spectra from program P223 were supplemented by two snapshots of {\hd},
one of which was obtained as an early-release observation (observation
X0240202; PI: J.~B.~Hutchings), while the other was obtained as part of a 
PI-Team  program to investigate the interstellar medium (observation P1030101;
PI: K.~R.~Sembach). 

Table~\ref{journal} provides details of the {\fuse} observations of {\hd}.  
Successive columns list the name of the data set; the heliocentric
Julian date (HJD) at the beginning of the integration;
the total integration time; the orbital phase at the start 
of the observation, which was computed from the ephemeris of \citet{Sterken97};
the mean flux and the standard deviation in a fiducial continuum band;
and the implied signal-to-noise ratio.
All the {\fuse} spectra of {\hd} were obtained through the large (LWRS) aperture.
Observation X0240202 was obtained in histogram (HIST) mode; all the others
were obtained in time-tag (TTAG) mode.

In addition to the {\fuse} spectra of {\hd}, a spectrum obtained on
1993 September 17 by the Berkeley Extreme and Far-UV Spectrometer (BEFS) 
on-board {\orfeus} was retrieved from the Multimission Archive at the Space 
Telescope Science Institute (MAST\footnote{The archiving of non-{\hst} data 
at MAST is supported by the NASA Office of Space Science via grant NAG5-7584 
and by other grants and contracts. STScI is operated by the Association of 
Universities for Research in Astronomy, Inc., under NASA contract NAS5-26555.}).
Details of this observation are also listed in Table~\ref{journal}.
This spectrum was obtained only $\sim$51 days before the eruption of
{\hd} began, and is the only UV spectrum that records the characteristics
of the system at this crucial time.
An additional stroke of luck is that the {\orfeus}/BEFS spectrum was obtained 
at $\phi=0.075$, i.e., just after primary eclipse, when {\stA} is ``in front" 
of its companion. 
Hence the  P~Cygni  absorption components in this spectrum reflect the wind 
conditions of the eruptor ({\stA}).  
However, the {\orfeus}/BEFS spectrum has poorer wavelength resolution 
and signal-to-noise (S/N) than the {\fuse} spectra.  
The spectrum obtained from MAST was not processed further, although 
a wavelength shift of approximately $+60$~{\kms} was applied in order to 
align interstellar features with their counterparts in the {\fuse} spectra.

Finally, as described below, our analysis of {\hd} was supplemented by 
{\fuse} spectra of two other stars in the SMC: 
{Sk\,80} \citep[AV\,232; O7~Iaf$+$; see][]{Crowther02} and 
{Sk\,108} \citep[AV\,332; WN4:$+$O6.5~I:; see][]{Mallouris03}.
These observations were obtained in TTAG mode through the LWRS (Sk\,80)
or MDRS (Sk\,108) aperture.
Additional details for these observations are also provided in 
Table~\ref{journal}.

\subsection{Processing of {\fuse} Spectra}\label{fuseproc}

The instruments on-board the {\fuse} satellite consist of four
aligned, prime-focus telescopes and Rowland circle spectrographs
\citep{Moos00,Sahnow00}.  
Two of these "channels" are coated with LiF to provide high efficiency
longward of $\sim$1000~{\AA}, while the other two are fabricated from
SiC to give high throughput between the Lyman limit and $\sim$1100~{\AA}. 
The spectra are recorded by two photon-counting detectors, each of which
consists of two segments.
Thus, a {\fuse} observation produces 8 independent spectra, which are
labelled by their channel (LiF1, LiF2, SiC1, SiC2) and their detector
segment (A or B).  
Except for a small region near 1100~{\AA}, the entire {\fuse} waveband is
redundantly covered by 2--4 of these spectra.

The {\fuse} spectra were uniformly extracted and calibrated with the
standard reduction pipeline (CalFUSE version 2.4.1).
Subsequent processing steps combined the 8 individual spectra from a given
observation into a single spectrum. 
Since the different spectra have inherently different quality (due to
the different effective areas of each channel) and suffer from instrument
artifacts to differing degrees (e.g., thermally induced channel misalignment),
these manipulations were conducted interactively in three steps.

First, residual zero-point shifts in the wavelength scales for the
different spectra were removed by comparing the observed positions of
H$_2$ lines formed in the interstellar medium of the Galaxy with the 
laboratory wavelengths listed by \citet{Jenkins97}.
The corrections derived in this way were applied to each spectrum individually.
The corrections for data from the SiC channels were always smaller than 
{0.25~\AA}, while for LiF data they were in general smaller than {0.03~\AA}.
The overlapping regions of spectra from different channels and detector
segments were checked carefully to ensure that the resultant wavelength
scales were consistent.  
When necessary, the spectra were trimmed to minimize the effect of nonlinearities 
in the wavelength scale due to residual distortion near the edges of detector
segments.

Second, the aligned spectra were inspected visually.  
In general, only the highest quality data for any particular wavelength
interval were retained for the merged spectrum.
Gaps in wavelength coverage were filled with data of poorer quality only
when necessary.
LiF2A spectra were favored over LiF1B spectra for coverage longward of
$\sim$1100~{\AA} because the flux calibration of LiF1B spectra is 
severely compromised by an optical artifact known as ``the worm" 
\citep{Sahnow03}.

Third, the aligned, trimmed, and hand-selected spectra from the various 
channels and detector segments were combined by using the 
{IRAF}\footnote{IRAF is distributed by the National Optical Astronomy 
Observatories, which are operated by the Association of Universities for 
Research in Astronomy, Inc., under cooperative agreement with the National 
Science Foundation.} routine {\tt scombine}.
The spectra contributing to the final, combined data product were assigned
equal weights (w $=$ 1) except for SiC1A spectra, which were given 
{w$=$0.25} because of their poorer quality.
Finally, the data were rebinned to a constant wavelength step of 0.05~\AA.

Seven of the eight P223 observations of {\hd} were combined to produce a 
global mean reference spectrum for the P223 data set. 
This average spectrum has S/N $\sim$40 per data sample in the continuum
near 1000~\AA. 
Observation P2230103 ($\phi=$0.124) was excluded because detector 1 was
not operating at full voltage for the entire exposure, so that spectra 
extracted from it (i.e., LiF1A, LiF1B, SiC1A, SiC1B) were anomalously noisy.

\subsection{Far-UV Continuum Fluxes}

Continuum placement in spectra of WR stars is very uncertain for all UV 
wavelengths due to the number and breadth of emission and P~Cygni wind profiles.
Shortward of $\sim$1000~{\AA}, the problem is aggravated by the large 
number of interstellar absorption lines, particularly from  H$_2$, and 
by the confluence of the Lyman series of atomic H. 
In addition, the spectrum of {\hd} contains absorption from a velocity 
system associated with the foreground supernova remnant 
SNR0057$-$7226.\footnote{In the {1200--1800~\AA} spectral range, the 
velocity system of the SNR is at $v_{helio} = +312 \pm 3$~{\kms} and
$+343 \pm 3$~{\kms}, and includes lines for ions ranging from
{\ion{O}{1}} to {\ion{C}{4}} \citep{Koenigsberger01}.
The same system has been detected in {\fuse} spectra of the
{\ion{O}{6}} resonance doublet \citep{Hoopes01,Danforth03}.}

Despite these complications, we identified several spectral bands that allow 
the continuum level to be defined reliably: e.g., 1010--1020~{\AA} and 
1055--1060~{\AA}.
In particular, the wavelength interval 1059.6--1060.6~{\AA} is
free from interstellar and photospheric lines.
We used the average value of the flux over this interval,
which we denote by {\fcont}, to characterize continuum variations of {\hd} 
in the far-UV.
The values of {\fcont}, along with the root-mean-square scatter over the 
continuum band [$\sigma$({\fcont})], and the corresponding S/N ratio are 
provided in Table~\ref{journal}.
\section{Far-Ultraviolet Light Curve of {\hd}}\label{fuselc}

The orbital period of {\hd} is 19.2654$\pm$0.0002 \citep{Sterken97}.
As indicated in Figure~\ref{orbit} (taken from Koenigsberger 2004), orbital phase {$\phi=0.0$} is defined 
by the minimum of the primary eclipse; i.e., when {\stA} eclipses {\stB}. 
Owing to the substantial eccentricity of the orbit 
\citep[$e = 0.32, 0.27, 0.30$ according to][respectively]{Breysacher91,Moffat98,
Kaufer02}, the secondary eclipse occurs at {$\phi=0.36$}.
The radius of {\stA} prior to eruption (R$_A(1978)=21$~\rsun) and the
radius of {\stB} (R$_B=15$~\rsun) were determined by 
\citet{Breysacher91} from analysis of the visual light curve.
Since the orbital inclination is $\sim 86\degr$
\citep{Breysacher91}, the visual continuum eclipse at $\phi=$0.0 is total.  
However, the current radius of {\stA} is uncertain.
In 1994 December, \citet{Koenigsberger98a} determined it to be 
{R$_A(1994)=35$~\rsun}.  
An analysis of the contemporary light curve is needed to confirm the
expectation that the radius of {\stA} has been decreasing systematically
over the intervening years.

Our {\fuse} observations permit a far-UV light curve to be derived, which
covers both eclipses.
The top panel of Figure~\ref{fuvlc} plots {\fcont} from Table~\ref{journal}
as a function of orbital phase.
The curve superposed on the data illustrates the predicted eclipse 
depths from the model described in \S\ref{model}.
The bottom panel of Fig.~\ref{fuvlc} shows the differential magnitude of 
{\hd} with respect to {Sk\,108} (large triangles), and also indicates 
the visual light curve determined by \citet[][~crosses]{Breysacher80}.
Although the number of far-UV data points is limited, there is generally
good agreement between the far-UV and optical light curves, {\em except}
for the {\fuse} data obtained in 1999 and 2000 (indicated by parentheses).
Evidently the system was systematically brighter in these early observations
compared with 2002, presumably because it was still declining from the maximum
brightness attained during its 1994 eruption.
The depths of the optical and far-UV eclipses at {$\phi=0.36$} are similar, 
while at {$\phi=0.00$ the far-UV eclipse seems to be deeper than the optical eclipse.
However, given the large optical photometric variations detected on timescales of
$\sim$6 hours, little more can be concluded from this single point at $\phi=$0. 

The {\orfeus}/BEFS measurement of the far-UV continuum flux is the point 
in Fig.~\ref{fuvlc} at {$\phi=0.075$}.
Although it has the largest uncertainty of any of the {\fcont} measurements, 
its mean value agrees very well with that obtained at nearly the same 
orbital phase from {\fuse} spectrum P2230102.
This agreement of pre- and post-outburst measurements separated by 8.9 years
(168 orbital cycles) is very interesting, since it suggests that {\hd} has
returned to its pre-outburst state.
In this context, it is interesting to note that even though the {\orfeus}/BEFS
spectrum was obtained shortly before the start of the eruption, the FUV
continuum fluxes did not presage dramatic changes.

\section{Far-UV Spectrum and Variability}\label{fusesp}

The average {\fuse} spectrum of {\hd} is illustrated in Figures~\ref{sp920} 
and \ref{sp1030}, where it is also compared with the spectrum of {Sk\,108} 
(dotted lines).  
{Sk\,108} is an ideal template for the analysis of {\hd} because: 
 1) it is also in the SMC and is thus similarly affected by reddening; and 
 2) it consists of a WN4 star and an O6.5-supergiant \citep{Foellmi03}, and 
    should resemble the combined spectrum of {\stB} $+$ {\stC}.
A striking difference between these objects is the number and intensity
of emission lines in the far-UV spectrum of {\hd}.  
While the {\fuse} spectrum of {Sk\,108} has strong emission only in
the {\ion{O}{6}} resonance doublet, the spectrum of {\hd} exhibits
emission features due to {\ion{S}{6}}, {\ion{N}{4}}, {\ion{N}{3}}, 
{\ion{He}{2}}, {\ion{O}{6}}, {\ion{S}{4}}, {\ion{P}{5}}, and {\ion{C}{4}}.

Assuming that the combined spectrum of {\stB} $+$ {\stC} resembles the 
spectrum of {Sk\,108}, then the source of the strong emission lines 
in {\hd} must be {\stA} and, possibly, the wind-wind collision (WWC) region.
At longer UV wavelengths (1200 -- 1700~\AA), most emission 
lines are also dominated by {\stA} \citep{Koenigsberger04}, although {\stB}  
appears to contribute significantly to the P~Cygni features at
{\ion{N}{5}~$\lambda$1240}, {\ion{C}{4}~$\lambda$1550}, and 
{\ion{He}{2}~$\lambda$1640}.
One feature that is associated exclusively with {\stA} (and, possibly, 
the WWC region) is the semi-forbidden {\ion{N}{4}]~$\lambda$1486} line, 
which displays extremely large line-profile variations.
These variations are usually interpreted in terms of a non-spherically 
symmetric line-emitting region.

In the {\fuse} wavelength region, the {\ion{P}{5}} {$\lambda\lambda$1117, 1128}
doublet presents a similar opportunity for isolating emission arising only in 
{\stA}, with the advantage that these resonance lines produce strong
P~Cygni absorption components.
For O-type stars, the intensity of the {\ion{P}{5}} doublet is correlated 
with the stellar luminosity \citep{Walborn02}, and the great strength of 
these lines in the spectrum of {\hd} is consistent with the 
large luminosity derived for {\stA} (\citet[][$L_A \sim 3 \times 10^6 L_\sun$]{Koenigsberger98b}).


We used the standard deviation of 7 of the 8 spectra obtained in 2002 
to quantify the spectral variability of {\hd}. 
As shown in Figs.~\ref{sp920} and \ref{sp1030}, the largest values of
this dispersion coincide with the positions of strong emission lines.
At most other wavelengths, the time variations have a base level of
$\sim0.5 \times 10^{-12}$ \funits}, which reflects the continuum 
variations due to the eclipses but also includes cycle-to-cycle fluctuations. 
Smaller variations occur only at wavelengths that coincide with saturated 
absorption lines; e.g., the strong absorptions
due to interstellar Lyman~$\beta$ at $\sim$1025~{\AA}.

The {\ion{P}{5} $\lambda$1117} line is only moderately affected by blending 
with other stellar and interstellar lines, and can thus be used 
to study line-profile variability in more detail.
Figure~\ref{p5montage} illustrates the orbital variations of 
{\ion{P}{5} $\lambda$1117} for the eight {\fuse} spectra obtained in 2002.
In this montage, the spectrum obtained near primary eclipse 
(P2230101; $\phi = 0.001$) is used as a template against which variations
in the other profiles are compared.
In order to make this comparison meaningful, two systematic effects were
allowed for.
First, the profiles were shifted in velocity so that they represent the 
emission in the frame of reference of {\stA}.
The radial velocity curve of \citet{Niemela97} was used to remove the
orbital motion of {\stA}.
Second, the profiles were re-normalized with respect to the sum of
the combined continua of {\stA} $+$ {\stB}. 
This was accomplished by removing the light of {\stC}, which was 
assumed to contribute approximately 33\% of the total continuum in the 
far-UV, as it does in the visual region of the spectrum \citep{Breysacher80}.
By taking the observed continuum level at phases 0.124 and 0.775 as 
representative of the sum of all three continua, the contribution from
{\stC} was estimated to be {$\sim2.1 \times 10^{-12}$ \funits} in the 
vicinity of the {\ion{P}{5}} doublet. 
This value was subtracted from each of the spectra in turn.  
Finally, the 8 {\ion{P}{5}} profiles were renormalized by dividing by 
the out-of-eclipse  continuum level of {\stA} $+$ {\stB} 
($4.2 \times 10^{-12}$ \funits). 

Fig.~\ref{p5montage} indicates that the primary cause of line-profile
variability in {\ion{P}{5}} is the appearance of strongly enhanced absorption 
(with respect to $\phi=0.001$) in the P~Cygni profile when {\stA} is 
``in front" of {\stB}.
Conversely, when {\stA} is eclipsed by {\stB} at $\phi \approx 0.360$,
the low-velocity part of the P~Cygni absorption trough is 
weakened compared with primary eclipse.
This clear pattern of behavior can be attributed to wind and stellar
eclipses, as demonstrated by the model discussed in the next section.
\section{Predicted Line Profile Variations from Eclipses}\label{model}
 
P~Cygni profiles in the spectrum of a hot star are produced in an extended and 
rapidly expanding stellar wind.  
Variability in these lines can be used to diagnose: 
 a) the presence of eclipses;  
 b) aspherical geometry of the wind due, e.g., to collisions with the
    outflow from a companion star; and  
 c) structural changes in the density, velocity, or ionization conditions
    in the wind.
Eclipses and wind-wind collisions produce periodic variability on orbital 
timescales.  
Wind eclipses, in particular, may be used to derive information about the 
geometry and structure of the stellar winds of the components of an early-type
binary system \citep{Munch50,Willis79,Koenigsberger90,Auer94}.

Figure~\ref{wwcgeom} provides a representation of the {{\stA} $+$ {\stB}} binary 
system at four orbital phases, under the assumption that both stars possess 
stellar winds.  
The geometry of the shock cone produced by the collision of these outflows 
is approximated with the formalism developed
by \citet{Canto96}, with stellar wind velocities estimated by 
\citet{Koenigsberger04}, 
{\vinf(\stA) $=$ 2000~\kms}, 
{\vinf(\stB) $=$ 2600~\kms} 
and estimates for the mass-loss rates of
{\mdot$_A = 1 \times 10^{-4}$~\msunpyr}, 
{\mdot$_B = 2 \times 10^{-5}$~\msunpyr}.
These mass-loss rates should be taken as representative values only, since the 
only reliable determination was made shortly after 
the eruption in 1994 \citep{Koenigsberger98b,Drissen01} and may be larger
than during quiescent times.
Furthermore, the wind velocity has varied significantly since the 1980s.  
We assume that the wind of {\stA} consists of an inner region 
extending out to a distance $r_{accel}$, within which the atmosphere 
accelerates and approaches the  maximum wind speed, {\vinf}.  
In Fig.~\ref{wwcgeom}, {$r_{accel} = 1.4$\,R$_A$} is represented by the 
circle traced by long dashes. 
Beyond $r_{accel}$, the wind expands at the constant speed {\vinf}.
The wind that flows in the direction of the companion undergoes
a highly supersonic shock when it encounters the companion's outflow,
which brakes the radial component of its motion and forces it to move
along the working surface of the shocks.
The kinematic properties of the shock surface are determined by the 
conservation of momentum in the hydrodynamical system; see, e.g.,
\citet{Canto96}.

Fig.~\ref{wwcgeom} illustrates that continuum radiation from {\stA} or {\stB}
can traverse several physically distinct regions before leaving the system.  
Consider, e.g., the elongation illustrated in Fig.~\ref{wwcgeom}(b). 
Starting from {\stB}, the short, dashed lines parallel to the 
line-of-sight cross in succession the wind of {\stB}, the shock cone surface, 
and finally the wind of {\stA} before reaching a distant observer located 
at the bottom of the figure.  
All these regions have velocity components in the direction of the observer,
and therefore all contribute to the shape of the line profile at negative 
velocities, i.e., to the morphology of the absorption trough of a 
P~Cygni profile from the binary system. 
Hence, while the emission component of the P~Cygni profile contains contributions 
from large volumes of wind material, the contribution to the absorption trough
is confined to the wind contained in the columns of material projected onto the 
stellar disks.

The discussion of wind-profile variability can be simplified if we choose a 
P~Cygni line that is formed in the wind of {\stA} but not in the wind of {\stB}.  
We argued above that the WN4:$+$O6.5~I: binary system {Sk\,108} should be a good 
template for the combined spectrum of {\stB} and {\stC}, since {\stB} 
is believed to be a WNE star.  
{Sk\,108} lacks P~Cygni emission due to {\ion{P}{5}} and thus we 
assume that the entire P~Cygni feature of {\ion{P}{5}} in {\hd} is
attributable to {\stA}.
In this case, we expect to observe the intrinsic P~Cygni absorption component 
of {\stA} at orbital phase $\phi=$0.00, when it fully eclipses its companion.  
At orbital phases immediately before and after $\phi=$0.00, the continuum 
radiation of {\stB} must traverse regions of the wind of {\stA} to reach the 
observer.  
This is the  ``wind eclipse". 
Because the wind of {\stA} has velocity components in the direction of {\stB} 
as well as in the direction of the observer, the wind eclipse causes an 
enhancement of the  P~Cygni absorption component and a reduction in the intensity 
of the emission component.  
The latter is due to absorption at positive velocities from the wind that is 
expanding toward {\stB}.  
At $\phi=$0.36, when {\stB} eclipses {\stA}, the P~Cygni absorption 
component should be significantly reduced, since the column of wind material
from {\stA}  where this absorption is produced is occulted by {\stB}.

\subsection{Model Calculation}

These concepts can be quantified by computing synthetic line profiles 
with a simple radiative transfer code that includes the effects of continuum 
and wind eclipses \citep{Auer94,Flores01}.
This program treats the radiative transfer using the Sobolev approximation only 
where this approximation is valid, and performs the exact integration of the 
radiative transfer equation in wind regions where the velocity gradient is too 
small to satisfy the Sobolev criterion.  
For simplicity, the  model assumes spherically symmetric 
line-forming regions and neglects both wind-wind collision effects and 
scattering processes.  
The model profiles thus provide baseline information on the eclipse effects 
alone.
Discrepancies between the observed and predicted line profile 
variations can be used to assess the importance of other phenomena not
incorporated in the model (e.g., the wind of {\stB}; wind-wind collisions; 
or an aspherical wind geometry).  

As before, we assume that the wind obeys a linear velocity law out to a distance 
$r_{accel}$, beyond which the velocity has the constant value {\vinf}.\footnote
{All distances are expressed in units of the radius of {\stA}, R$_A$. All
velocities are in units of the ``turbulent velocity", \vturb.}
In addition, we assume that the ionization structure is such that {\ion{P}{5}} 
is present to a distance $r_{max}$. 
The computer program allows for the possibility of a radial dependence of the 
ionization structure through the parameter $f(r)$, which is given by 
$f(r)=\chi(r) v(r) r^2$, where $\chi$(r) is the opacity.  
Through a trial-and-error process, the free parameters of the model were 
adjusted until the general features of the {\ion{P}{5} $\lambda$1117} line 
profile at $\phi=$0.001 were reproduced. 
Emphasis was placed on trying to match the shape of the P~Cygni absorption 
component.  
This was achieved with the set of input parameters labeled ``Mp01-1"
in Table~\ref{modelpar}.
Figure~\ref{p5mp1} compares the synthetic line profile computed with these
parameters for $\phi=0.001$ (solid line) with the observed profile.

A satisfactory match between the synthetic and observed {\ion{P}{5}} 
P~Cygni profile of {\stA} could only be achieved with a very fast wind
velocity law.
Table~\ref{modelpar} shows that the terminal speed, $v_{max}$, of 
model Mp01-1 is achieved within $r_{accel} = 1.4$\,R$_A$.
The consequences of varying $r_{accel}$ are illustrated in the profiles
plotted at the bottom of Fig.~\ref{p5mp1}:
a slower velocity (corresponding to $r_{accel} = 1.6$\,R$_A$) produces a P~Cygni
absorption trough that is too deep at low velocities (short dashes),
while a faster velocity law (corresponding to $r_{accel} = 1.3$\,R$_A$)
produces too little low-velocity absorption (long dashes).
Since rapid acceleration (i.e., a ``fast" velocity law) is a key feature
of winds driven by radiation pressure in spectral lines, these results suggest
that the current wind of {\stA} is driven by this mechanism.


For a fixed opacity structure, the maximum extent of the line-emitting region, 
$r_{max}$, determines the intensity of the line profile, since a larger volume 
contains a larger number of line-emitting ions.   
The consequences of varying $r_{max}$ are illustrated by the profiles in
the upper section of Fig.~\ref{p5mp1}.
Since the long- and short-dashed curves bracket the plausible range of
fits to the observed profile, we conclude that 
{3.3\,R$_A$ $\leq r_{max} \leq $ 4.5\,R$_A$} for the 
region emitting {\ion{P}{5} $\lambda$1117} radiation.
%

The  value of the terminal wind speed, {\vinf $=$ 1750~\kms}, constrains 
the product $(v_{max}) \times (v_{turb})$.  
However, small values of {\vturb} produce very sharp and deep absorption 
at {\vinf}, contrary to what is observed.
Instead, we find that values of {\vturb} of $\sim$70--80~{\kms}
(which corresponds to $\sim$5\% of \vinf) are required to match the 
soft blue edge of the absorption trough.
As a result, the ``edge" velocity (i.e., the velocity where the blue edge of the
absorption trough meets the continuum) is significantly faster than {\vinf}.
This is a characteristic of strong P~Cygni absorption troughs in O-type
stars in general \citep{Prinja90}.
The occurrence of soft blue absorption edges is generally attributed to the 
presence of additional velocity dispersion in the wind, which is frequently 
characterized as ``turbulence".
However, despite its widespread occurrence, the origin of this velocity
dispersion is not known.

The other significant discrepancy between the model and the observed line 
profiles is the sharp emission peak near line center; see Fig.~\ref{p5mp1}.
The red component of the {\ion{P}{5}} doublet at 1127~{\AA} provides a
built-in check of this discrepancy. 
Since the doublet separation ($+$2693~\kms) is greater than {\vinf} for 
{\stA} (1660 \kms; see Table~\ref{windpar}), the two components are 
not radiatively coupled and can be treated independently. 
The oscillator strength of red component of the {\ion{P}{5}} doublet is
half that of the blue component, so the model profile for 
{\ion{P}{5} $\lambda$1127} was calculated with half the opacity, $f(r)$,
listed for Mp01-1 in Table~\ref{modelpar}, but holding all other parameters
fixed.
Figure~\ref{p5doub} shows the synthetic profile for the doublet obtained
by adding the computations for each component together.
It confirms that the synthetic profiles provide a good representation of the
observed line strengths, and that the sharp peak near line center is the 
most discrepant feature for {\em both} components of the doublet.
(Note that the strong absorption at $\sim$1000~{\kms} in the
absorption trough of the red component is an interstellar feature.)
We will show below that this problem may be solved if we assume that 
{\stC} is responsible for the excess emission near 0 velocity.


The model parameters listed in Table~\ref{modelpar} also permit line
profiles to be computed for different orbital phases.
In the model, phases are characterized in terms of the impact parameter,
$p$, which is defined as the minimum distance between the 
line-of-sight to {\stB} from the center of {\stA}.
For illustrative purposes, $p$ can be read directly from the abscissae 
in Fig.~\ref{orbit}. 
The phases for the model computations were chosen to have impact parameters
corresponding to the phases of the {\fuse} observations.


The phase-resolved progression of synthetic line profiles is shown in 
Figure~\ref{phasemp1}, where each profile is labeled with the 
corresponding impact parameter (expressed in units of R$_A$).
The line profile computed for primary eclipse ($p=0$) is used as template,
to show the predicted range over which excess absorption should be 
present.
The left panel corresponds to orbital phases when {\stB} is occulted by 
{\stA}, while the right panel shows the  phases when
the portions of the wind of {\stA} are eclipsed by {\stB}.
The profiles in both panels can be compared directly with their counterparts
in Fig.~\ref{p5montage}.

Fig.~\ref{phasemp1} shows that the P~Cygni absorption trough of {\ion{P}{5}}
in the spectrum of {\stA} should decrease in strength near secondary eclipse,
while the emission lobe is relatively unaffected.  
Thus, the line profile behavior shown in Fig.~\ref{p5montage}
is explained qualitatively by the combined effects of wind and physical eclipses.  
There is, however, one important feature of the theoretical prediction that 
does not appear in the observations: for small impact parameters 
(e.g., $p \approx 2$\,R$_A$), the model predicts enhanced absorption over the 
{\em entire} velocity range from {$-$\vinf} to {$+$\vinf}, projected onto {\stB}.
In contrast, the observations only show excess absorption from {$-$\vinf} 
to $ v \approx 0$.  

Thus, it appears that either:
 (a) the wind of {\stA} flowing toward {\stB} does not have the same 
     structure as the wind emerging from the opposite side of {\stA}; or
 (b) the predicted excess absorption on the ``red" side of the line profile
     is filled in by emission from some process that is not
     currently included in the model. 
For (b), this emission must arise from gas that is not projected against 
either {\stA} or {\stB}, and which is receding from the observer at orbital phases 
$\sim 0.0 \pm 0.15$.
A plausible candidate that would produce emission is the WWC region
illustrated in Fig.~\ref{wwcgeom}(a). However, the opening angle of the
shock cone would need to be sufficiently small for the projection of the
WWC flow velocities along line-of-sight to be as large as $+$1000~{\kms}.
In addition, this putative emission would have to be visible at the opposite 
orbital phases ($\phi\sim 0.36$) at similar speeds approaching the observer.   
Although the observed profile at this orbital phase does display what could be
interpreted as excess emission, the same feature can also be interpreted in 
terms of a {\em physical} eclipse of the absorption-forming column of material 
in the wind of {\stA}.
This effect is predicted by the model: see, e.g., the $p=0.3$~R$_A$ profile
in the right-hand panel of Fig.~\ref{phasemp1}.
Thus, it seems more likely that the wind emitted by opposing hemispheres 
of {\stA} has substantially different structure.

\subsection{Correcting for Line Emission From {\stC}\label{emC}}

If {\stC} is a mid-O type supergiant \citep{Koenigsberger02}, its spectrum 
should contain prominent P~Cygni profiles in the resonance lines of 
{\ion{P}{5}}; see, e.g., \citet{Walborn02}.  
We used an archival {\fuse} spectrum of {Sk\,80} (O7~Iaf$+$) as a template
for {\stC}, and subtracted it from each spectrum of {\hd} in order to
compensate for the contamination it introduces.
Since the continuum level of {Sk\,80} matches the estimated continuum
strength of {\stC} (i.e., $\sim$33\% of the total recorded for {\hd}), this
operation also corrects the {\fuse} spectra for the contribution to the
continuum from {\stC}.
Figure~\ref{p5montmSk80} illustrates the 8 line profiles from {\fuse} program
P223 after subtracting the spectrum of {Sk\,80} and correcting for the
orbital motion of {\stA} (as in Fig.~\ref{p5montage}).

A new set of models was computed for the {\ion{P}{5}~$\lambda$1117} profile
of {\stA} at $\phi=0.001$.
Matching parameters to achieve a good fit with the observed profile yielded 
once again $r_{accel}=1.4$\,R$_A$, but required larger values of
$r_{max}$ and $f(r)$.
  The adopted input parameters are listed in Table~\ref{modelpar} under the label
``Mp01-2".
The increase in $r_{max}$ and $f(r)$ compensate for the fact that
the P~Cygni profile of the {\ion{P}{5}} resonance line in the spectrum of 
{Sk\,80} has a prominent, low-velocity absorption component.
Subtracting this profile from the spectrum of {\hd} produces more emission,
and consequently the model requires more optical depth over a larger volume 
to achieve a good match.

Figure~\ref{p5mp2} shows the {\ion{P}{5}~$\lambda$1117} profile
of {\stA} at $\phi=0.001$ after the removal of the contribution from {\stC}.
The ``best-fit" profile computed with the parameters of model Mp01-2
is shown as a continuous line, which corresponds to $r_{max}=5.3$.
For comparison, models with  $r_{max}=5.6$~R$_A$ and $r_{max}=4.8$~ R$_A$ are 
also shown.  
The match is significantly better than the one illustrated in Fig.~\ref{p5mp1},
especially in the emission lobe near $v=0$. 
Thus we conclude that this feature is indeed an artifact of 
stationary ``third light" contamination by {\stC}.

It is important to note that the assumed  {\ion{P}{5} contribution from {\stC} is
overestimated  by adopting Sk 80 as a template.  This line's emission is 
 maximum in supergiants at O6$-$O7 (Walborn et al. 2002), while {\stC} is likely
to be O4$-$O6.  Thus, {\stA}'s actual wind structure is between that given by
models Mp01-1 and Mp02-2, although the  uncertainty does not affect the conclusions
regarding the wind accelerating region. Sk 80 is the ``earliest" O-supergiant 
in the SMC for which there exists a FUV spectrum, and thus provides the only
template we can use at this time. 

The phase-dependent line profile variability predicted by this second set of
calculations is shown in Figure~\ref{phasemp2}.
The variations are qualitatively similar to those generated from model Mp01-1;
compare, e.g., with Fig.~\ref{phasemp1}.
The main difference is the more pronounced wind eclipse due to the larger 
physical extent of the {\ion{P}{5}} emitting region which, in this case, 
extends beyond the orbital radius.


\subsection{Absolute Dimensions of the System}

A theoretical light curve can be constructed from the model continuum
intensities that were required to achieve a good match with the observed 
line profile of {\ion{P}{5} $\lambda$1117}.
An additional, unknown requirement is the size of {\stB} relative to {\stA}.
By varying this single parameter until a good fit with the
far-UV light curve (Fig.~\ref{fuvlc}) was achieved, we determined that
R$_B = 0.7$\,R$_A$ in 2002.
By further adopting R$_B=$15 R$_\odot$ \citep{Breysacher91}, we infer that
the far-UV continuum of {\stA} was R$_A=$21\,R$_\odot$ in 2002.
With this value of R$_A$, the length scale for all model parameters is 
established, which in turn allows the distances in Figures~\ref{orbit} and
\ref{wwcgeom} to be calibrated in units of R$_\sun$.
Table~\ref{windpar} lists the derived parameters for the wind of {\stA} in 2002.


It is interesting to note that the value of R$_A$ that we find from the 
{\fuse} light curve obtained in 2002 is the same as that deduced by 
\citet{Breysacher91} from optical data obtained in 1978.
This coincidence suggests that the optically thick radius marking the base of 
the stellar wind of {\stA} has not changed, despite the disruptions caused by
the eruptive phenomenon that took place in the system.


\section{Discussion and Conclusions}\label{concl}

\subsection{The Wind Structure of Star~A}

We have used phase-resolved wind profiles of the {\ion{P}{5}} resonance 
doublet obtained by {\fuse} in 2002 together with a simple wind eclipse
model to constrain the velocity law of {\stA}, the relative continuum
intensities of {\stA} and {\stB}, and the physical dimensions of the
{\hd} system.
We find that, to a first approximation, the {\ion{P}{5}} wind profiles
require that the wind of {\stA} is expanding with a ``fast" velocity
law, as is typical of outflows driven by radiation pressure in spectral lines.
Systematic variations of profile strength and morphology as a function of 
orbital phase are well reproduced by a combination of ``wind eclipses" of 
{\stB} and physical eclipses of {\stA} by {\stB}.

However, there are four important discrepancies between the observations and the 
synthetic line profiles obtained from our simplified model.   
The first consists of the presence of an emission ``spike" 
near the rest velocity in the components of the {\ion{P}{5}} resonance doublet
that is not predicted by the model.
We show in \S\ref{emC} that this problem can be resolved by assuming that the
excess emission arises from the third component of the system, {\stC}.
The presence of a {\ion{P}{5}} wind feature is consistent with the classification
of {\stC} as an O4--O6 supergiant on the basis of its photospheric absorption lines 
\citep{Koenigsberger01}.
Unfortunately, solutions for the three remaining discrepancies are less 
straightforward.


\noindent{\bf Absence of a wind eclipse at $v \geq 0:$}
The wind eclipse around phase $\sim$0.0 produces reduced
emission over the range of velocities associated with the column of material 
projected against the disk of {\stB}.
We have shown that the wind of {\stA} expands rapidly, so that it 
achieves its terminal speed before it reaches the radius
of its relative orbit about {\stB}.
Thus, at phases immediately before and after $\phi=$0.00, the entire 
emission lobe should be significantly reduced in intensity by
the wind eclipse. 
However, the observations show the expected effect only at 
$v \leq 0$~{\kms}; i.e., only on the blue side of the P~Cygni profile.
  
Thus, it seems as if only the portion of the wind of {\stA} that is 
approaching the observer has the geometry and kinematics assumed by 
the model.
We see two possible explanations for this discrepancy: 
\begin{enumerate} 
  \item
    There is an extended region of {\ion{P}{5}} emission associated with 
   {\stB} that compensates for the absorption produced by the wind eclipse. 
    This emission could come from the surface of the shock cone associated
    with the WWC region, which effectively truncates the spherical symmetry 
    of the outflow.
   
  \item    
    The wind of {\stA} is not spherically symmetric, but is 
    significantly perturbed in the direction of {\stB}.  
    However, the asymmetry must be present in the wind acceleration region, 
    which we have shown lies very close to the surface of {\stA}, far from 
    the expected location of the WWC stagnation point.  
    Hence, if this explanation is correct, we conclude that the 
    wind structure of {\stA} in the direction of its companion 
    differs from that of other directions, as opposed to a
    wind structure that is truncated by the WWC region.  This difference could
    either be {\em intrinsic} to {\stA} or could be the result of ``sudden radiative
    braking" \citep{Gayley97}.  
    Note that if the wind structure towards the companion is non-standard, 
    then the WWC surfaces that are drawn in Fig.~\ref{wwcgeom} are not valid.  
    For example, a slower wind towards the companion would have the effect 
    of moving the stagnation point towards {\stA}.

\end{enumerate} 

\noindent{\bf Strength of P~Cygni Absorption at $\phi=$0.36:}
If {\ion{P}{5} $\lambda$1117} is formed exclusively in the wind of {\stA}, 
then the eclipse model predicts that the absorption trough of its P~Cygni 
profile should exhibit significant weakening during secondary eclipse, because {\stB} occults the column of material 
responsible for producing this absorption.
However, only a modest degree of weakening is observed at slow velocities;
see, e.g., Fig~\ref{p5montage} and Fig.~\ref{p5montmSk80}.
At large velocities, $v \geq v_\infty$, the absorption seems to be even
more pronounced during secondary eclipse.
A natural explanation for this behavior is that the wind of {\stB}
{\em also emits and absorbs} {\ion{P}{5}}; i.e., that our assumption
concerning the unique association of {\ion{P}{5}} with {\stA} is not
correct.
Unfortunately, the present version of the wind-eclipse model is not
able to test this idea, since a rigorous calculation must include 
detailed radiative transfer through two stellar winds and the WWC region.
Instead, we simply note that if the spectrum of {\stB} does posses a 
P~Cygni wind profile in the {\ion{P}{5}} resonance doublet, then it
differs significantly from the spectrum of the WR star in {Sk\,108}.

\noindent{\bf Extent of the P~Cygni absorption edge:}
The final discrepancy is the extent of the P~Cygni absorption  edge.  
Although {\vinf $=$ 1750~\kms} for {\stA}, the edge velocity is 
$v_{edge} \geq -2200$~{\kms}.  
The greater extent of $v_{edge}$ is a well-known phenomenon in O-type and 
WR stars, so it is not too surprising to find it in {\hd} as well.  
What is surprising, however, is that the extended, ``soft" blue edge of
the P~Cygni absorption trough does not vary with orbital phase.
Since {\stB} also drives a strong stellar wind, it is possible that its
radiation field further accelerates the outflow from {\stA} that
has already achieved {\vinf}.
However, since this speculation cannot be tested directly with the present
data set, the issue must remain open for now.

\subsection{New Insights into the Eruption Mechanism}

The changes in the wind of {\hd} prior to the eruptions in 
1993--1994 were characterized by a progressive decrease in
terminal velocity, which was accompanied by a systematic increase in 
density \citep{Koenigsberger98b}.
The first sign that a significant change was occurring in {\hd} can be
found in its {\iue} spectra of 1986 \citep{Koenigsberger04}, although
peculiarities were already present in the early 1980s \citep{Niemela88}.  
The amplitude of the perturbation seems to have grown gradually over the
subsequent $\sim$13 years until some kind of critical state was reached.
This critical state produced the sudden eruption in 1993, which was
followed by a second, stronger eruption in 1994.
According to the visual data of Albert Jones 
\citep[reproduced in][]{Koenigsberger04}, the eruption started around 
HJD 2,449,299 (1993 November 7), which is only 51 days (2.6 orbital cycles) 
after the {\orfeus}/BEFS spectrum was acquired.


Figure~\ref{cforfeus} compares the {\ion{P}{5} $\lambda$1117} wind profile
in the {\orfeus}/BEFS spectrum with its counterpart in the {\fuse} P2230102 
spectrum.  
Both spectra were obtained at essentially the same orbital phase, though they
are separated by 168 orbital periods. 
In addition to slightly stronger P~Cygni absorption around 
$-500$~{\kms}, the only morphological difference between the two spectra 
is the location of the discrete absorption component that marks the position 
of terminal speed: 
in 1993 {\vinf $=$ 1530~\kms}, while {\vinf $=$ 1750~\kms} in 2002.
Hence, the trend for decreasing wind velocity persisted from the early 1980s 
up to $\sim$51 days before the start of the eruption.


Evidently, the overall structure of the wind of {\stA} was very similar
just prior to the first eruption in 1993 and in 2002, when {\hd}
was in a quiescent state.
Hence, we infer that the wind must have been driven by the same mechanism 
at both epochs; and we have argued on the basis of the rapid acceleration
required to model the {\ion{P}{5}} wind profile that in 2002, this mechanism 
was radiation pressure in spectral lines.
Consequently, in 1993 the conditions in the star must have changed from
supporting a ``normal" radiatively driven outflow to a state of significantly
enhanced mass loss in less that 51 days; i.e.,
whatever set of critical conditions initiated the eruption 
occurred within a remarkably short time interval.
This rapid reconfiguration of the properties of the wind provides a strong
constraint on the unknown mechanism responsible for the eruption.

The fact that the wind velocity decreased systematically for many years
before the eruption suggests that some critical limit was reached in the
autumn of 1993.
If the decreasing wind velocity is associated with an increasing wind density, 
as appears to be the case from the growth of emission line intensities over the 
1980--2000 timescale \citep{Koenigsberger04}, it is possible that the 
critical conditions are related to the 
``bi-stability limit" \citep{Lamers95,Vink99}.
In this case, the eruptions would be interpreted as the observable
manifestation of the transition from the ``fast, rarefied wind" side of 
the bi-stability limit to the ``slow, dense wind" side and the concomitant 
re-structuring of the circumbinary environment.
Since two major eruptions occurred within $\sim$1.5 years, it appears that
{\stA} remained very close to its ``bi-stability limit" for this long
before beginning to relax to its pre-outburst state.

Of course, the crucial unanswered question is what underlying evolutionary
process caused {\stA} to move towards its ``bi-stability limit" in the first
place, e.g., by triggering enhanced mass-loss rate or decreasing
the velocity of the radiatively driven outflow.
Identifying this trigger mechanism in {\hd} is certain to improve our 
understanding
of eruptive phenomena that occur in other massive stars and binary systems.
Since these outbursts may drive significantly greater mass loss than possible 
via stellar winds, they have great potential to alter both the evolutionary
histories of these stars and the yields of chemically enriched 
material they provide to their local environments.

\acknowledgements
GK thanks Albert Jones for providing his visual magnitude estimates.  
This investigation was supported by PAPIIT/DGAPA grants 118202, 119205 and by
CONACYT grant 36569.

\bibliographystyle{aj}

%
%
\begin{figure}
     \plotone{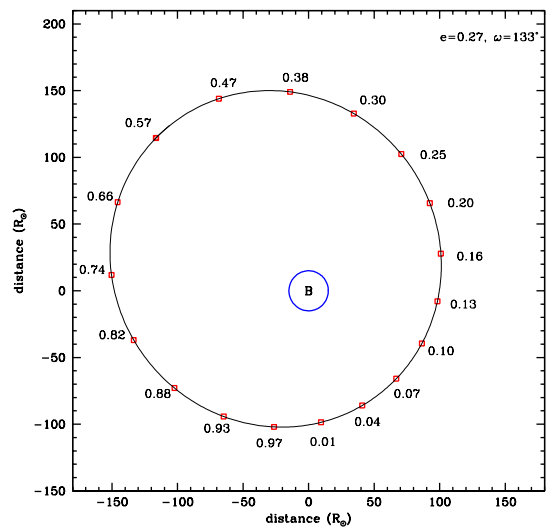}
     \figcaption{
       Schematic representation (from Koenigsberger 2004) of the orbit of {\stA} in a reference 
       frame that is fixed with respect to {\stB}, viewed from above
       the orbital plane.  
       The observer views the system from a perspective that is nearly in 
       the plane of the page, looking``up" from the bottom 
       along a line through impact parameter (abscissa) 0.
       The open squares are located at 20$\degr$ intervals and are
       labelled by their orbital phases. 
       \label{orbit}
      }
\end{figure}
%
\begin{figure}
     \plotone{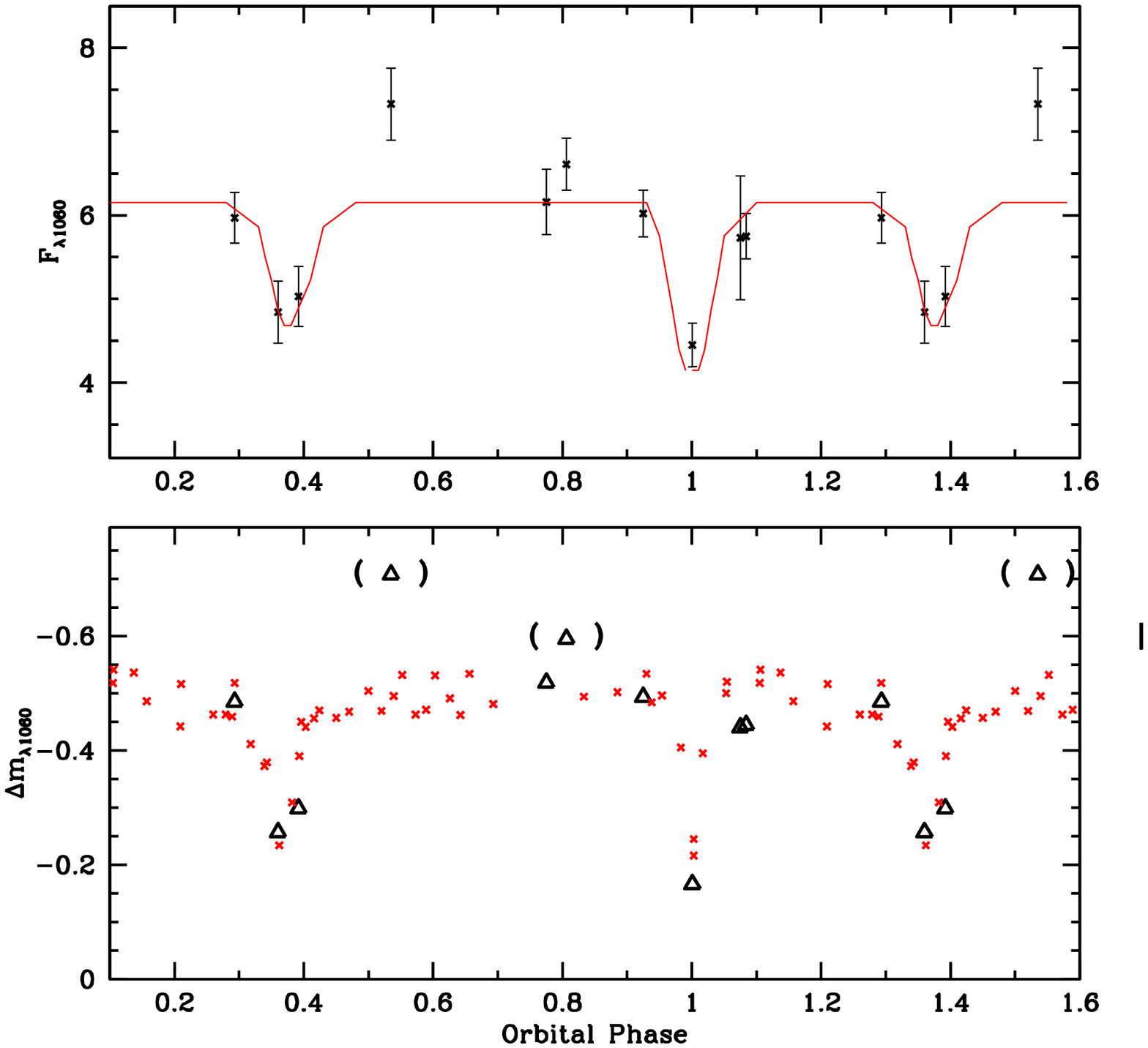}
     \figcaption{
      Far-UV continuum light curve of {\hd}.
      Top: the average continuum flux {\fcont}, expressed in units of 
      {$10^{-12}$~\funits} and not corrected for reddening.
      Error bars indicate $\sigma$({\fcont}); see Table~\ref{journal}.
      The solid curve corresponds to the  prediction of the model described 
      in \S\ref{model}.
      Bottom: the magnitude difference between {\hd} and {Sk\,108} (triangles) 
      compared with the visual light curve in 1978
      \citep[crosses;~][]{Breysacher80}.
      {\fuse} measurements from 1999 and 2000 are enclosed in parentheses.
      Evidentally {\hd} was systematically brighter in the far-UV at these 
      epochs.
      \label{fuvlc}
      }
\end{figure}
%
\begin{figure}
     \epsscale{0.90}
     \plotone{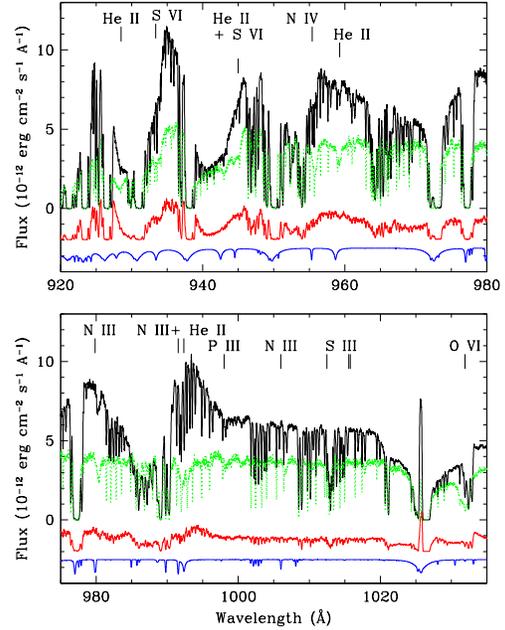}
     \figcaption{
      The short-wavelength region of the mean far-UV spectrum of {\hd} in 2002. 
      The light tracing indicates the mean spectrum of {\hd}; see
      \S\ref{fuseproc}.
      The dark lines show the standard deviation about the mean spectrum
      caused by orbital variations.
      This ``variation spectrum" has been
      arbitrarily shifted by {$-2\times10^{-12}$ \funits}.
      The dotted line shows the {\fuse} spectrum of the SMC binary {Sk\,108}
      (WN4:$+$O6.5~I:), while a normalized absorption-line spectrum from a 
      Kurucz LTE, line-blanketed model atmosphere is plotted at the bottom
      (shifted by{$-2.5\times10^{-12}$ \funits}). 
      Major features are identified.
      \label{sp920}
      }
\end{figure}
%
\begin{figure}
     \epsscale{0.90}
     \plotone{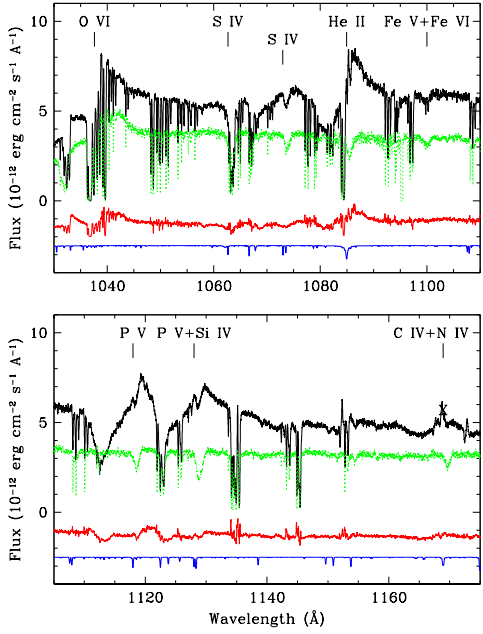}
     \figcaption{
      Same as Fig.~\ref{sp920} for the long-wavelength region of
      the {\fuse} waveband.
      The emission feature marked by an ``X" is due to scattered
      solar light.
      \label{sp1030}
      }
\end{figure}
%
\begin{figure}
     \plotone{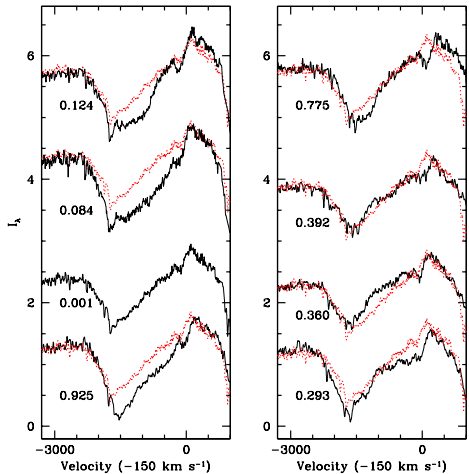}
     \caption{
     Line profiles of {\ion{P}{5} $\lambda$1117} at different orbital phases, 
     compared to the line profile at $\phi=$0.001 (dotted lines).   
     The wind eclipse (left panel) only affects approaching velocities.
     \label{p5montage}
     }
\end{figure}
%
\begin{figure}
     \plotone{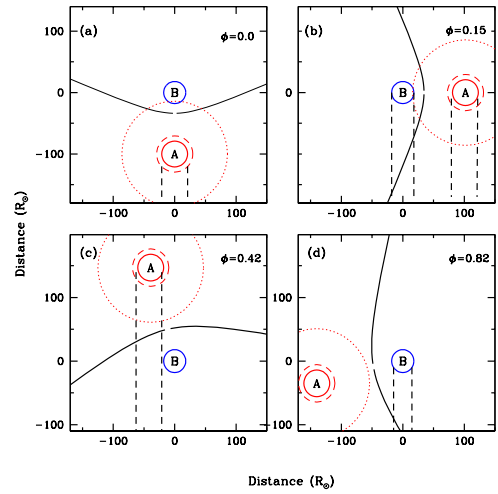}
     \figcaption{
     Geometry of the orbit and WWC region for different phases of the
     {\hd} system.
     The observer is at the bottom in the four panels.
     Vertical dashes indicate the columns of wind material that produce 
     P~Cygni absorption troughs, which includes regions where enhanced
     absorption is produced due to wind eclipses.
     The circular broken lines around {\stA} show the extent of
     the accelerating wind (1.4\,R$_A$).  
     The dotted circles show $r=4.1$\,R$_A$, which marks the estimated
     lower limit of the {\ion{P}{5}} line emitting region; see \S\ref{model}.
     \label{wwcgeom}
     }
\end{figure}
%
\begin{figure}
     \plotone{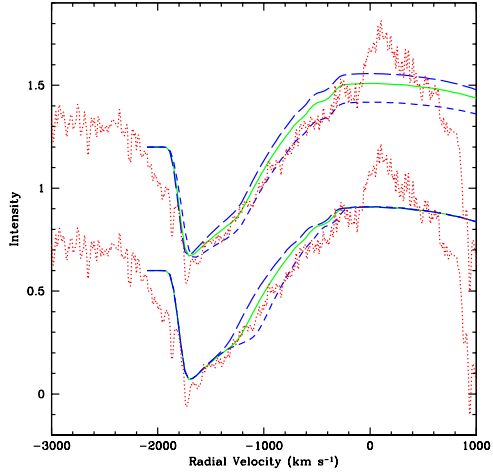}
     \figcaption{
      The observed line profile of {\ion{P}{5}~$\lambda$1117} at 
      $\phi=0.001$ (dotted) compared with the profile synthesized from
      model Mp01-1 (light solid).
      Discontinuous lines show computed line profiles with all but one
      model parameter held constant.
      Top: Effect of varying the maximum extent of the line-emitting
      region.  
      Short dashes: $r_{max} = 3.3$~R$_A$; 
      long dashes: $r_{max} = 4.5$~R$_A$.
      Bottom: Effect of varying the wind velocity law.
      Short dashes: $r_{accel} = 1.6$~R$_A$; 
      long dashes: $r_{accel} = 1.3$~R$_A$.
      \label{p5mp1}
      }
\end{figure}
%
\begin{figure}
     \plotone{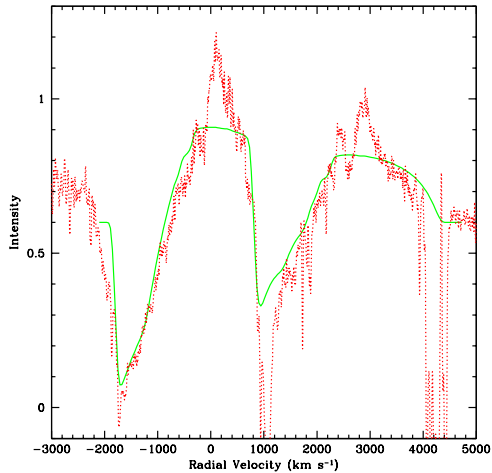}
     \figcaption{
      Comparison between observed and synthesized profiles of the 
      {\ion{P}{5} $\lambda\lambda$1117, 1128} doublet at $\phi=$0.001.
      Excess emission is apparent near the center of both lines.      
      \label{p5doub}
      }
\end{figure}
%
\begin{figure}
     \plotone{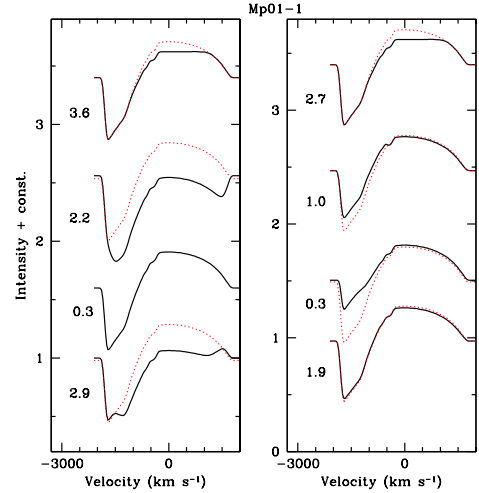}
     \figcaption{
      Orbital line-profile variations computed from model Mp01-1.
      Individual profiles are labeled the value of the impact parameter 
      (expressed in units of R$_A$) of the line-of-sight to {\stB} from 
      the center of {\stA}, and correspond to the orbital phases covered
      by the {\fuse} observations (Fig.~\ref{p5montage}).
      \label{phasemp1} 
      }
\end{figure}
%
\begin{figure}
    \plotone{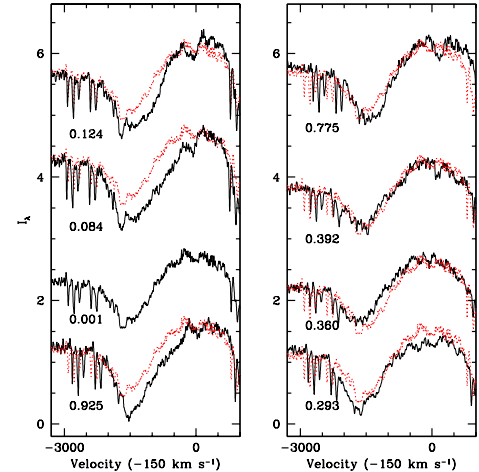}
    \figcaption{
     Line profiles of {\ion{P}{5} $\lambda$1117} at different orbital phases, 
     compared profile at $\phi=$0.001 (dotted lines).
     In contrast to Fig.~\ref{p5montage}, the spectrum of {Sk\,80} has been
     subtracted to account for the contributions of {\stC}.
     The sharp emission excess seen near $v=0$ in Fig.~\ref{p5montage}
     is no longer present.
     \label{p5montmSk80}
     }
\end{figure}
%
\begin{figure}
     \plotone{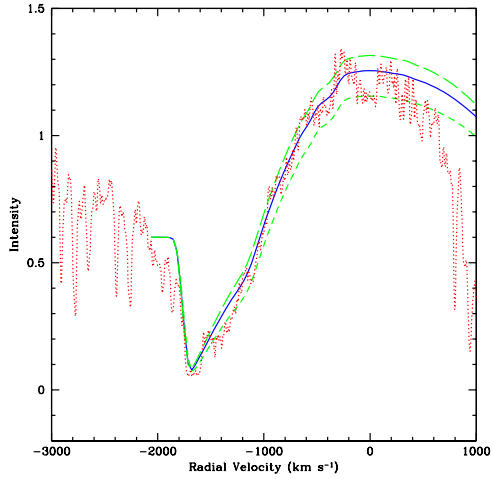}
     \figcaption{
      Model fit to the {\ion{P}{5} $\lambda$1117} profile after subtraction
      of the spectrum of {Sk\,80}, which was used as a proxy for the
      contributions of {\stC}.
      The continuous line is the closest match to the data, which was
       computed with the parameters of model Mp01-2 (Table~\ref{modelpar}).
      The long and short dashes correspond to the same model but with 
      $r_{max}=5.6$~R$_A$ and $r_{max}=4.8$~R$_A$, respectively.
      \label{p5mp2}
      }
\end{figure}
%
\begin{figure}
     \plotone{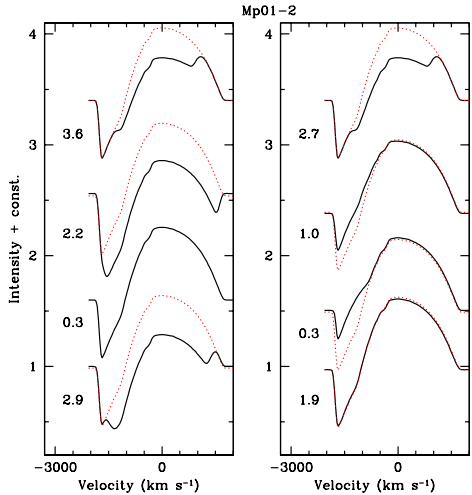}
     \figcaption{
      Orbital line-profile variations computed from model Mp01-2.
      \label{phasemp2}
      }
\end{figure}
%
\begin{figure}
    \plotone{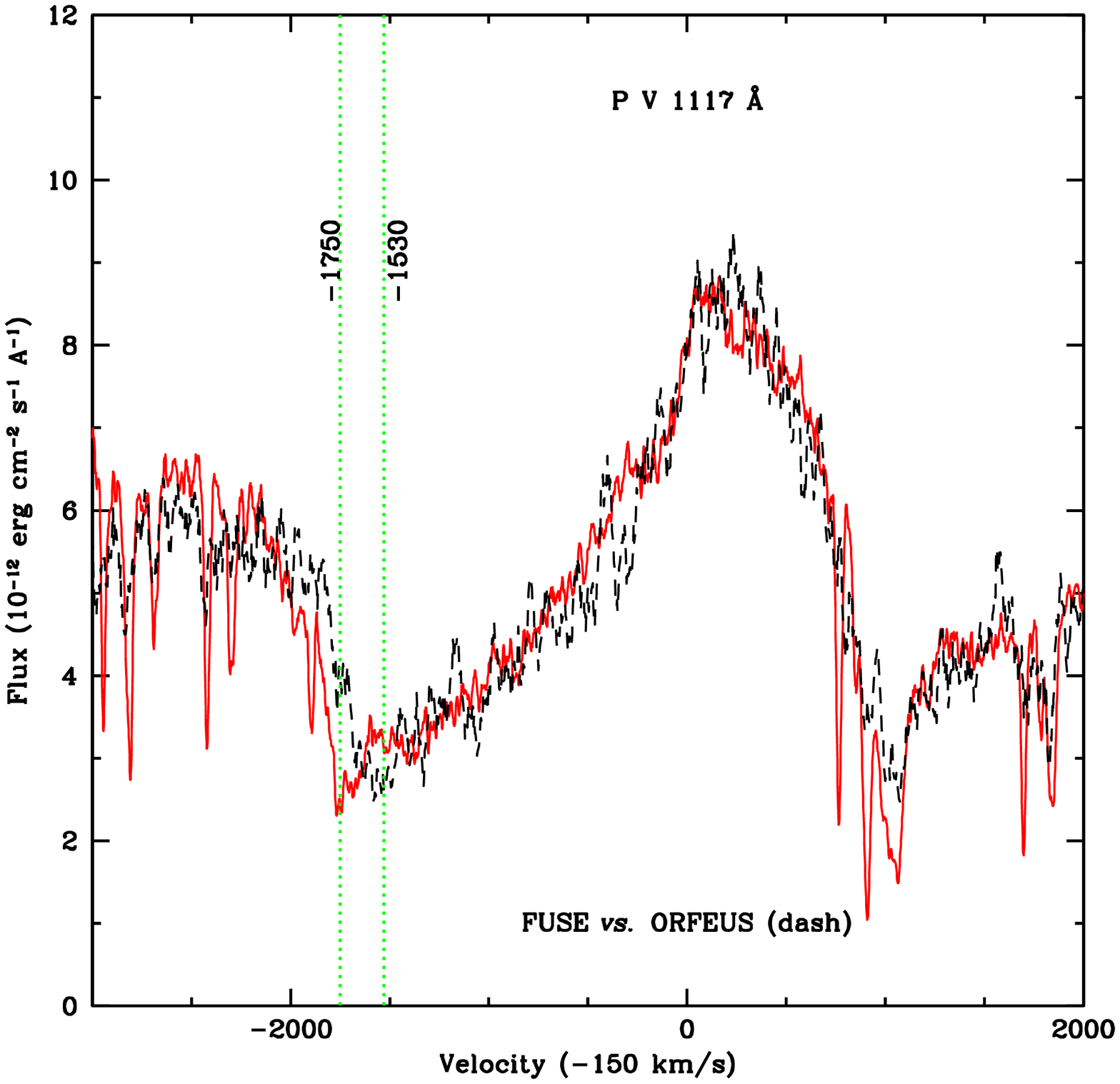}
    \figcaption{
     Comparison between P~Cygni profiles of the {\ion{P}{5} $\lambda$1117}
     resonance line observed by {\orfeus}/BEFS in 1993 (dashed line;
     $\phi=0.075$) and {\fuse} in 2002 (solid line; $\phi=0.084$).
     The {\orfeus}/BEFS spectrum was shifted in velocity by 
     $-65$~{\kms} to align nearby interstellar lines and by 
     {$+0.5 \times10^{-12}$ \funits} to align the continuum levels. 
     The velocity of the discrete absorption component at the time of the 
     {\orfeus} observation was 220~{\kms} slower than at the time of 
     the {\fuse} observation, but otherwise the profiles are very 
     similar.
     \label{cforfeus}
     }
\end{figure}
\newpage
\input{tab1.tex}    
\input{tab2.tex}    
\input{tab3.tex}    
\end{document}

%% file: tab1.tex
\begin{deluxetable}{llccccc}
\tablecaption{Journal of Observations\label{journal}}
\tablecolumns{7}
\tablewidth{0pt}
\tablehead{ \colhead{Observation}                         & 
            \colhead{HJD\tablenotemark{a}}                &
	    \colhead{Int. Time [ks]}                      &
            \colhead{Phase\tablenotemark{b}}              &
            \colhead{{\fcont}\tablenotemark{c}}           &
            \colhead{$\sigma$({\fcont})\tablenotemark{d}} &
            \colhead{S/N}                                 }
\startdata
\sidehead{\hd}
BEFS1068\tablenotemark{e}          
         & 49248.0358 & \phn1.810 & 0.075    & 5.73:                 &  0.74:  & \phn7   \\
X0240202 & 51472.3964 & \phn3.153 & 0.535    & 7.33                  &  0.43   & 17      \\
P1030101 & 51728.0736 & \phn5.745 & 0.806    & 6.61\tablenotemark{f} &  0.31   & 21      \\
P2230105 & 52432.2891 & \phn4.342 & 0.360    & 4.84                  &  0.37   & 13      \\
P2230107 & 52478.8300 & \phn3.953 & 0.775    & 6.16\tablenotemark{f} &  0.39   & 16      \\
P2230108 & 52481.7160 & \phn8.050 & 0.925    & 6.02                  &  0.28   & 22      \\ 
P2230101 & 52483.1789 & \phn6.957 & 0.001    & 4.45                  &  0.26   & 17      \\
P2230102 & 52484.7828 & \phn6.925 & 0.084    & 5.75                  &  0.27   & 21      \\
P2230103 & 52485.5381 & \phn7.894 & 0.124    & \nodata               & \nodata & \nodata \\
P2230104 & 52488.8000 & \phn7.871 & 0.293    & 5.97                  &  0.30   & 20      \\
P2230106 & 52548.5025 & \phn3.886 & 0.392    & 5.03                  &  0.36   & 14      \\
\sidehead{Sk\,80}	  
P1030201 & 51728.2801 &    11.700 & \nodata  & 2.25                  &  0.10   & 23      \\ 
\sidehead{Sk\,108}
P1030401 & 51728.5619 &    13.958 & \nodata  & 3.82                  &  0.14   & 27      \\
\enddata
\tablenotetext{a}{HJD $-$ 2,400,000.0 at the start of the integration.}
\tablenotetext{b}{Orbital phase computed from the ephemeris of \cite{Sterken97}.}
\tablenotetext{c}{Mean continuum flux between 1059.6 and 1060.6~{\AA}, except for Sk 80 where
                  the interval 1057$-$1058~{\AA} was used,  
                  in units of $10^{-12}$ {\funits}.}.
\tablenotetext{d}{Standard deviation in the mean continuum flux.}
\tablenotetext{e}{Observation with {\orfeus}/BEFS.}
\tablenotetext{f}{There is a broad dip at 1058.54~{\AA} at this phase.}		  
\end{deluxetable}

%% file: tab2.tex
\begin{deluxetable}{lccl}     
\tablecaption{Parameters for \ion{P}{5} $\lambda$1117 Model Calculation\label{modelpar}}  
\tablecolumns{4}
\tablewidth{0pt}
\tablehead{ \colhead{Parameter}   & 
            \colhead{Mp01-1}      &
	    \colhead{Mp01-2}      & 
	    \colhead{Description}  }
\startdata
$r_{max}$    &  4.1      &  5.3      & Maximum extent of emitting region           \\
\vinf        &  25.0     &  20.8     & Terminal speed of the wind                  \\
\vturb       &  70       &  80       & ``Turbulent" speed in \kms                  \\
$r_{accel}$  &  1.4      &  1.4      & Radius over which the wind is accelerating  \\
$f_0$        &  400      &  400      & Opacity factor at the stellar surface       \\
$f(r), r_1$  & 2600, 1.3 & 4100, 1.3 & Opacity factor out to $r_1$                 \\
$f(r), r_2$  &  220, 4.1 &  220, 5.3 & Opacity factor out to $r_2$                 \\
I$_A$, I$_B$ &   48, 32  &   48, 32  & Continuum intensity of each star            \\
$r_{orb}$    &   4.76    &   4.76    & Mean orbital separation                     \\ 
R$_B$        &    0.7    &   0.7     & Radius of {\stB} in units of R$_A$          \\
\enddata 
\end{deluxetable}

%% file: tab3.tex
%
\begin{deluxetable}{lrll}
\tablecaption{Wind Parameters of Star A in 2002\label{windpar}}
\tablecolumns{4}
\tablewidth{0pt}
\tablehead{ \colhead{Parameter}           & 
            \multicolumn{2}{c}{Value}     & 
	    \colhead{Comments}            }
\startdata
$R_{max}$   &  111 & \rsun &  Maximum extent of emitting region $=r_{max}$ \vturb \\
$V_\infty$  & 1660 & \kms  &  Terminal speed of the wind of {\stA}, $=$\vinf \vturb \\
\vturb      &   80 & \kms  & ``Turbulent" speed in the wind in {\stA}                      \\
$R_{accel}$ &   29 & \rsun &  Radius of accelerating wind region for {\stA}, $=r_{accel} R_A$ \\
R$_{A}$     &   21 & \rsun &  Radius of {\stA} in 2002                                  \\
\enddata
\end{deluxetable}